\documentclass[aps,pra,superscriptaddress,twocolumn,floatfix,a4paper]{revtex4}

\usepackage{graphicx,graphics,epsfig}   
\usepackage{dcolumn}    
\usepackage{bm}         
\usepackage{amsmath}    
\usepackage{verbatim}   
\usepackage{color}      
\usepackage{subfigure}  
\usepackage{times,natbib}
\usepackage{amsmath,amsfonts,amssymb,graphics,graphics,color,times}

\usepackage{latexsym}
\usepackage{amsmath}
\usepackage{amssymb}
\usepackage{amsfonts}
\usepackage{amsthm}
\usepackage{mathrsfs}
\usepackage{color,verbatim,graphics}
\usepackage{psfrag}
\DeclareMathAlphabet{\mathrsfs}{U}{rsfs}{m}{n}
\DeclareMathAlphabet{\mathpzc}{OT1}{pzc}{m}{it}
\DeclareMathAlphabet{\matheus}{U}{eus}{m}{n}
\DeclareMathAlphabet{\mathbbold}{U}{bbold}{m}{n}

\def\one{\leavevmode\hbox{\small1\normalsize\kern-.33em1}}
\def\tr{\mbox{tr}}

\newcommand{\HH}{\mathcal{H}}

\newcommand{\CC}{\mathbb{C}}

\newcommand{\ba}{\begin{eqnarray}}
\newcommand{\ea}{\end{eqnarray}}
\newcommand{\ban}{\begin{eqnarray*}}
\newcommand{\ean}{\end{eqnarray*}}

\newcommand{\ket}[1]{|#1\rangle}

\begin{document}

\title{Device-independent tomography of multipartite quantum states}

\author{K\'aroly F. P\'al}
\affiliation{Institute for Nuclear Research, Hungarian Academy of Sciences, H-4001 Debrecen, P.O. Box 51, Hungary}
\author{Tam\'as V\'ertesi}
\affiliation{Institute for Nuclear Research, Hungarian Academy of Sciences, H-4001 Debrecen, P.O. Box 51, Hungary}
\author{Miguel Navascu\'es}
\affiliation{Universitat Aut\`onoma de Barcelona, 08193 Bellaterra (Barcelona), Spain}

\begin{abstract}
In the usual tomography of multipartite entangled quantum states
one assumes that the measurement devices used in the laboratory
are under perfect control of the experimenter. In this paper,
using the so-called SWAP concept introduced recently, we show how
one can remove this assumption in realistic experimental
conditions and nevertheless be able to characterize the produced
multipartite state based only on observed statistics. Such a black
box tomography of quantum states is termed self-testing. As a
function of the magnitude of the Bell violation, we are able to
self-test emblematic multipartite quantum states such as the
three-qubit W state, the three- and four-qubit
Greenberger-Horne-Zeilinger states, and the four-qubit linear
cluster state.
\end{abstract}

\maketitle

\section{Introduction}\label{intro}

Quantum entanglement \cite{rmp_horo} plays a prominent role in
quantum theory, and particularly in quantum information. Indeed, a
big effort has been devoted to its characterization and detection
recently \cite{review_toth}.

In usual tomography of entangled quantum states, one has to rely
on certain assumptions about the measurement devices used in the
experiment. These assumptions are usually difficult to meet in
practice. For instance, the characterization of a quantum state
cannot be considered conclusive if the devices
implementing the specific measurement operators are not under
precise control of the experimenter \cite{rosset}.

In the last years, the experimental preparation of complex
multipartite states has become a routine. State of the art
photonic experiments can generate and characterize six-qubit
entangled states \cite{photonic,pistate}. More recently, 14
entangled qubits were generated in ion-trap experiments
\cite{monz,lanyon}. Such is the range of qubits, for which, in
order to do a full tomography and reconstruct completely the
produced multipartite state, one has to resort to additional
information about the state. Such additional knowledge has been
exploited in the literature for states of low rank~\cite{gross},
for a matrix product state~\cite{mps} or for a permutationally
invariant (PI) state~\cite{pistate}. Although these extra
assumptions may simplify the analysis considerably, the
characterization of the quantum state usually becomes less
accurate.

In this paper, we follow a different approach based on the
so-called device-independent paradigm (see \cite{review_scarani}
for a review), which regards the local systems as black boxes with
some input and outputs and is minimalist in the sense that it
requires only the no-signalling assumption and that inputs are
freely chosen.

Tomography of quantum states in this device-independent framework,
where one characterizes multipartite states based only on lists of
statistical data coming from a Bell-type experiment, was termed
self-testing in the seminal work of Mayers and Yao~\cite{my}. At
that time the task of self-testing was mostly applied in the ideal
situation (see pioneering works in Refs.~\cite{selftest_ideal} as
well). Later, this limitation has been removed and since then a
number of works \cite{selftest_nonideal} have demonstrated
self-testing robust to external noise. However, the noise to be
tolerated in these schemes was extremely small. A resolution to
this issue was given by Ref.~\cite{swap}, which could extend
self-testing of quantum states and measurement devices to
realistic experimental situations. As an illustration of the power
of the so-called SWAP method of \cite{swap},
it has been proved in the bipartite case that a CHSH \cite{chsh} violation of $2.57$
certifies a singlet fidelity of more than $70\%$.

In this paper, making use of the SWAP method, we move from the
bipartite to the multipartite domain by self-testing famous
multipartite states such as the W state \cite{dur}, the
$3\&4$-qubit Greenberger-Horne-Zeilinger (GHZ) states \cite{ghz},
and the 4-qubit cluster state \cite{cluster} (recall that each of
these states has been implemented in the lab in photonic
experiments about a decade
ago~\cite{w_exp},\cite{ghz_exp},\cite{cluster_exp}). Note that in
our task of self-testing we do not assume any knowledge regarding
the specific workings of the experimental devices (such as the
dimension of the underlying Hilbert spaces or the type of
measurements involved), however, we accept that quantum theory
holds exactly.

To this end, we introduce the framework of Bell nonlocality tests.
Consider three distant observers, Alice, Bob, and Cecil, and allow
each of them to choose freely between two ($i=1,2$) dichotomic
observables, $A_i=\pm 1$, $B_i=\pm 1$, and $C_i=\pm 1$,
respectively. In a specific run of the experiment, the
correlations between the observations can be represented by the
product of the type $A_iB_jC_k$. The correlation function is then
the average over many runs of the experiment $\langle
A_iB_jC_k\rangle$ for $i,j,k=0,1,2$ (where we have chosen
$A_0=B_0=C_0=1$ to account for subcorrelation terms). In quantum
mechanics, the above mean value can be calculated as follows:
\begin{equation}
\langle A_iB_jC_k\rangle =
\tr\left(\rho\cdot \hat{A_i}\otimes\hat{B_j}\otimes
\hat{C_k}\right),
\end{equation}
where $\rho$ denotes Alice, Bob and Cecil's tripartite state, and
we have set $\hat{A_0}=\hat{B_0}=\hat{C_0}=\one$.

Note: we never use the fact that the underlying black box
state is pure. And we shouldn't, because, in that case, we just
have to show correlation in order to prove entanglement. We do
assume, however, that measurements are projective.

Remarkably, there exist situations in this setting where the observed statistics $\{\langle
A_iB_jC_k\rangle\}_{i,j,k}$ suffice to determine the underlying state
$|\bar{\psi}\rangle$ and observables $\bar{A}_i,\bar{B}_j,\bar{C}_k$, up to
local isometries and some additional (but
irrelevant) degrees of freedom. For instance, let us consider the
following famous set of correlations
\begin{align}\label{ghzextremal}
&\langle A_1B_1C_1\rangle = 1,\nonumber\\
&\langle A_1B_2C_2\rangle = \langle A_2B_1C_2\rangle = \langle
A_2B_2C_1\rangle = -1,
\end{align}

\noindent exhibiting the so-called GHZ
paradox~\cite{ghz},\cite{merminpara}. It has been shown recently
that the only state compatible with these correlations is the
famous GHZ state (up to local isometries and adding local
ancillary systems to the state)~\cite{mckague}. However,
in realistic experimental conditions, we cannot hope that the
above averages attain $\pm 1$ exactly. In order to quantify how
close the actual state in the box $\rho\in B(\HH_{box})$ is to our
mathematical guess $\ket{\bar{\psi}}\in \CC^d$, we must hence
introduce a figure of merit. A quite significant one is the
fidelity modulo local isometries, defined as

\begin{equation}
F=\max_{U}\langle\bar{\psi}|\tr_{junk}(U\rho U^\dagger)|\bar{\psi}\rangle.
\end{equation}

\noindent Here the ``junk'' system denotes extra degrees
of freedom which are not necessary -in first approximation- to
capture the physics of the experiment, and the maximization is
performed over all local isometries $U:\HH_{box}\to \CC^d\otimes
\HH_{junk}$.

Our task is to estimate the minimal value of the fidelity
$F$ compatible with the observed statistics $\{\langle
A_iB_jC_k\rangle\}_{i,j,k}$ (note that $F=1$ with respect to some
reference state $\ket{\bar{\psi}}$ implies perfect self-testing).
For didactic purposes, in this work we will not discuss
self-testing criteria which require the knowledge of the whole set
of correlations. Rather, we will investigate how the fidelity $F$
with respect to multipartite (three-qubit and four-qubit) states
varies as a function of the magnitude of violation of specific
Bell inequalities. This will be possible thanks to the recently
developed SWAP method~\cite{swap}.

Let us mention some recent works in the spirit of our paper, where
information regarding the state produced could be extracted from
multipartite Bell experiments: In Ref.~\cite{gme}, genuine
multipartite entanglement could be detected from Bell-type
inequalities, which test was implemented experimentally as well
recently~\cite{gmeexp}. Another promising method was proposed by
Moroder et al.~\cite{moroder}, which method provides access to certain
properties of a composite system via Bell inequalities, such as
negativity \cite{neg} and can be extended to the multipartite
realm (see also \cite{structure} for related results).
Finally, we would like to call the attention of the reader
to the very much related work of \cite{mirror_paper}, where, also via the SWAP tool,
the authors manage to derive a new Bell inequality to self-test the W state.

The paper is structured as follows. First, in Section~\ref{PI}, we
introduce our main tool, multipartite permutationally invariant
(PI) Bell inequalities, i.e., those which do not change under
exchanging parties. In Section~\ref{idea} we sketch the idea of
constructing PI Bell inequalities which are maximally violated by
PI states such as Dicke states. In Section~\ref{Wviol}, for
clarity of presentation, the method is introduced through the
example of the three-qubit W state (one of the simplest Dicke
states). In this way, we derive a couple of candidate Bell
inequalities for self-testing of W states.
Section~\ref{swapmethod} utilizes the SWAP method~\cite{swap} to
certify minimal fidelity with respect to the W state as a function
of violation of our Bell inequalities. This is done in
Sec.~\ref{selftestW}. Using known Bell inequalities from the
literature, we also self-test the (three-qubit and four-qubit) GHZ
states in Sec.~\ref{selftestGHZ} and the four-qubit cluster state
in Sec.~\ref{selftestcluster}. Section~\ref{conc} ends with a
conclusion, where we also pose some open questions.

\section{Tools}\label{tools}

\subsection{Permutationally Invariant Bell inequalities}\label{PI}

Bell-type inequalities are the central tool of our investigations
\cite{bell}. We shall focus on multipartite Bell polynomials which
are permutationally invariant, that is, they are symmetric under
any permutation of the parties. Each observer can choose between two
possible measurements featuring binary outputs. We use the
following simplified notation to represent such a PI Bell
inequality:
\begin{align}\label{sym}
\left[ \alpha_1 \mbox{ } \alpha_2 ; \mbox{ } \alpha_{11} \mbox{ } \alpha_{12} \mbox{ } \alpha_{22} \right]  \equiv&  \alpha_1(A_1 + B_1) + \alpha_2 (A_2 + B_2)  \nonumber  \\
        &+ \alpha_{11}A_1B_1 + \alpha_{12}(A_1B_2 + A_2B_1) \nonumber\\ & +
        \alpha_{22}A_2B_2,
\end{align}
where $A_i=\pm 1$ denotes the outcome of Alice's measurement
settings $i=1,2$. Likewise for Bob's settings. The extension to
more parties is straightforward. For instance, for $N=3$ parties,
the Mermin inequality \cite{mermin}, usually written as
\begin{equation}
M_3 = A_1B_1C_1 - A_1B_2C_2 - A_2B_1C_2 - A_2B_2C_1\leq 2
\end{equation}
now reads
\begin{equation}\label{mermin3} M_3 =  \left[ 0 \mbox{ }0  \mbox{ }; 0\mbox{ } 0\mbox{ } 0\mbox{ } ; 1\mbox{ } 0 \mbox{ }-1\mbox{ }0 \right] \leq 2.
\end{equation}
Here the maximum algebraic sum of $M_3=4$, corresponding to the
set of correlations~(\ref{ghzextremal}), is attained with a
three-qubit GHZ state~\cite{ghz}:
\begin{equation}
\label{ghzstate3} GHZ_3=(|000\rangle + |111\rangle)/\sqrt 2.
\end{equation}
and Pauli $\hat X$ and $\hat Y$ measurements.

Let us turn to the case of 4 parties. The generalized
Mermin-Ardehali-Belinskii-Klyshko \cite{MABK} (MABK) Bell
inequality for $N=4$ is given by
\begin{equation}\label{mermin4}
M_4 =  \left[ 0 \mbox{ }0  \mbox{ } ; 0\mbox{ } 0\mbox{ } 0\mbox{
} ; 0\mbox{ } 0 \mbox{ }0\mbox{ }0 ; 1\mbox{ } 1 \mbox{ } -1\mbox{
} -1 \mbox{ } 1 \right] \leq 4.
\end{equation}
Here the quantum maximum reads $8\sqrt 2$, which can be obtained
by using $\hat X$ and $\hat Y$ Pauli measurements and a four-qubit
GHZ state~\cite{ghz}:
\begin{equation}
\label{ghzstate4} GHZ_4=(|0000\rangle + |1111\rangle)/\sqrt 2.
\end{equation}

\subsection{Basic idea of our method}\label{idea}

Our aim is to create Bell inequalities which are
maximally violated by a given $N$-qubit PI state. The existence of
such Bell inequalities is a necessary condition for self-testing
of PI states. For simplicity, we focus on permutationally
invariant Bell inequalities with two measurements per
party~\cite{PI_Bell}, moreover we restrict ourselves to orthogonal
measurement settings lying in the $X-Z$ plane. These kind of
settings are tailored to the SWAP method \cite{swap} which will be
used in section~\ref{swapmethod} for the purpose of self-testing.

Let us now give a short description of our linear programming
based method focusing on the W state (but we believe that the
procedure can be generalized to any PI state, such as Dicke states
\cite{dicke}). Given our desired W state and orthogonal
measurement settings, we construct the Bell operator (with yet
unknown coefficients) and derive conditions for the Bell
coefficients to guarantee that the W state is an eigenstate of
this Bell operator. By our specific measurement angles in the
$X-Z$ plane, we next derive further conditions which ensure that
the Bell value (i.e. the mean value of the Bell operator with the
W state) does not change in first order on small variations around
these measurement angles. Finally, we enforce (linear) constraints
to bound the local value of the Bell expression, and maximize the
quantum value. The problem to be solved is one of linear
programming. We can further put extra constraints in this linear
program to find Bell inequalities which have a special structure
(e.g., which have no single party marginal terms). Let us stress
that the conditions we impose are not necessarily sufficient to
guarantee the optimality of the W state for getting maximal Bell
violation. However, in practice, it works well. In the next
section, we give a detailed description of this method.

\subsection{Illustration of the method via the W state}\label{Wviol}

In the case of PI Bell inequalities with two binary settings per
party, there are nine independent Bell coefficients and we can
write the Bell inequality in the notation of section~\ref{PI} as:
\begin{equation}
{\cal B} = \left[ b_1 \mbox{ }b_2  \mbox{ } ; b_3\mbox{ } b_4\mbox{ } b_5\mbox{ } ; b_6\mbox{ } b_7 \mbox{ }b_8\mbox{ }b_9\right] \leq L,
\label{eq:Bell_ineq}
\end{equation}
where $L$ is the local maximum.

Our aim is to construct a Bell inequality which is maximally
violated by the 3-qubit W state \cite{dur} given as:
\begin{equation}
|W\rangle\equiv\frac{1}{\sqrt
3}(|001\rangle+|010\rangle+|100\rangle). \label{eq:Wstate}
\end{equation}
The operators of the measurements we have taken are the same for
each party, that is $\hat A_1=\hat B_1=\hat C_1\equiv \hat M_1$
and $\hat A_2=\hat B_2=\hat C_2\equiv \hat M_2$. With this choice,
the Bell operator may be written as
\begin{equation}
\hat{\cal B}=\sum_{i=1}^9 b_i\hat G_i, \label{eq:Bell-operator}
\end{equation}
where
\begin{align}
\hat G_1\equiv& \hat M_1 \one \one + \one \hat M_1 \one + \one \one \hat M_1\nonumber\\
\hat G_2\equiv& \hat M_2 \one \one + \one \hat M_2 \one + \one \one \hat M_2\nonumber\\
\hat G_3\equiv& \hat M_1 \hat M_1 \one + \hat M_1 \one \hat M_1 + \one \hat M_1 \hat M_1\nonumber\\
\hat G_4\equiv& \hat M_1 \hat M_2 \one + \hat M_2 \hat M_1 \one + \hat M_1 \one \hat M_2\nonumber\\
& + \hat M_2 \one \hat M_1 + \one \hat M_1 \hat M_2 + \one \hat M_2 \hat M_1\nonumber\\
\hat G_5\equiv& \hat M_2\hat M_2 \one+\hat M_2 \one\hat M_2+\one\hat M_2\hat M_2\nonumber\\
\hat G_6\equiv& \hat M_1\hat M_1\hat M_1\nonumber\\
\hat G_7\equiv& \hat M_1\hat M_1\hat M_2+\hat M_1\hat M_2\hat M_1+\hat M_2\hat M_1\hat M_1\nonumber\\
\hat G_8\equiv& \hat M_1\hat M_2\hat M_2+\hat M_2\hat M_1\hat M_2+\hat M_2\hat M_2\hat M_1\nonumber\\
\hat G_9\equiv& \hat M_2\hat M_2\hat M_2. \label{eq:B_operators}
\end{align}
Note above we used the shorthand $\hat M_i\hat M_j\hat M_k$ for
denoting the tensor product $\hat M_i\otimes \hat M_j\otimes \hat
M_k$. If there are only two binary measurements per party, the
maximum violation can always be achieved with measurements
performed on qubits in the $X-Z$ plane (real qubits). The
corresponding measurement operators are linear combinations of the
Pauli operators $\hat X$ and $\hat Z$:
\begin{align}
\hat M_1&=\cos\varphi_1 \hat Z+\sin\varphi_1 \hat X,\nonumber\\
\hat M_2&=\cos\varphi_2 \hat Z+\sin\varphi_2 \hat X.
\label{eq:M12phi12}
\end{align}
Then it follows from Eqs.\
(\ref{eq:Bell-operator}-\ref{eq:M12phi12}) that the Bell operator
may also be expressed as
\begin{equation}
\hat{\cal B}=\sum_{i=1}^9\eta_i\hat H_i,
\label{eq:Bell-operatorH}
\end{equation}
where
\begin{align}
\hat H_1\equiv& \hat Z\one\one+\one \hat Z\one+\one\one \hat Z\nonumber\\
\hat H_2\equiv& \hat X\one\one+\one \hat X\one+\one\one \hat X\nonumber\\
\hat H_3\equiv& \hat Z\hat Z\one+\hat Z\one \hat Z+\one \hat Z\hat Z\nonumber\\
\hat H_4\equiv& \hat Z\hat X\one+\hat X\hat Z\one+\hat Z\one \hat X\nonumber\\
&+\hat X\one \hat Z+\one \hat Z\hat X+\one \hat X\hat Z\nonumber\\
\hat H_5\equiv& \hat X\hat X\one+\hat X\one \hat X+\one \hat X\hat X\nonumber\\
\hat H_6\equiv& \hat Z\hat Z\hat Z\nonumber\\
\hat H_7\equiv& \hat Z\hat Z\hat X+\hat Z\hat X\hat Z+\hat X\hat Z\hat Z\nonumber\\
\hat H_8\equiv& \hat Z\hat X\hat X+\hat X\hat Z\hat X+\hat X\hat X\hat Z\nonumber\\
\hat H_9\equiv& \hat X\hat X\hat X. \label{eq:H_operators}
\end{align}
The $\eta_i$ coefficients will depend on the choice of the
measurement operators, that is the choice of the measurement
angles $\varphi_1$ and $\varphi_2$. The state giving the maximum
quantum violation is the eigenstate belonging to the largest
eigenvalue of the Bell-operator with the measurements chosen
optimally. Therefore, we must make sure that the W state is an
eigenstate of the Bell operator, that is $\langle\psi|\hat{\cal
B}|W\rangle=0$ for all states $|\psi\rangle$ orthogonal to
$|W\rangle$. From $\hat Z|0\rangle=|0\rangle$, $\hat
Z|1\rangle=-|1\rangle$, $\hat X|0\rangle=|1\rangle$ and $\hat
X|1\rangle=|0\rangle$ it is not difficult to derive:
\begin{align}
\hat H_1|W\rangle&=|W\rangle\nonumber\\
\hat H_2|W\rangle&=2|\bar W\rangle+{\sqrt 3}|000\rangle\nonumber\\
\hat H_3|W\rangle&=-|W\rangle\nonumber\\
\hat H_4|W\rangle&=2{\sqrt 3}|000\rangle\nonumber\\
\hat H_5|W\rangle&=2|W\rangle+{\sqrt 3}|111\rangle\nonumber\\
\hat H_6|W\rangle&=-|W\rangle\nonumber\\
\hat H_7|W\rangle&=-2|\bar W\rangle+{\sqrt 3}|000\rangle\nonumber\\
\hat H_8|W\rangle&=2|W\rangle-{\sqrt 3}|111\rangle\nonumber\\
\hat H_9|W\rangle&=|\bar W\rangle,
\label{eq:HW}
\end{align}
where
\begin{equation}
|\bar W\rangle\equiv|D^2_3\rangle=\frac{1}{\sqrt
3}(|011\rangle+|101\rangle+|110\rangle). \label{eq:Wbarstate}
\end{equation}
From Eqs.\ (\ref{eq:Bell-operatorH}) and (\ref{eq:HW}) it follows that $|W\rangle$ is an eigenstate
of $\hat{\cal B}$ if:
\begin{align}
2\eta_2-2\eta_7+\eta_9&=0\nonumber\\
\eta_2+2\eta_4+\eta_7&=0\nonumber\\
\eta_5-\eta_8&=0
\label{eq:eigenstate}
\end{align}
The first, second and third lines follow from the requirements that $\langle\bar W|\hat{\cal B}|W\rangle=0$,
$\langle 000|\hat{\cal B}|W\rangle=0$ and $\langle 111|\hat{\cal B}|W\rangle=0$, respectively. The
expectation value of $\hat{\cal B}$ is:
\begin{equation}
q\equiv\langle W|\hat{\cal
B}|W\rangle=\eta_1-\eta_3+2\eta_5-\eta_6+2\eta_8.
\label{eq:expect}
\end{equation}

Another requirement to be ensured is that the measurement
operators chosen are optimal. For that it is necessary that the
maximum eigenvalue of $\hat{\cal B}$ remains unchanged due to
infinitesimal variations of $\varphi_1$ and $\varphi_2$. If
$|W\rangle$ is the appropriate eigenvector, the derivatives of
$\langle W|\hat{\cal B}|W\rangle$ in terms of these angles have to
be zero (the change of the eigenvector due to the variations of
the angles gives only second order contributions). Let us specify
the measurements to be orthogonal to each other, that is
$\varphi_1=\varphi$ and $\varphi_2=\varphi-\pi/2$. In the
Appendix~\ref{app} we show that the following new extra condition
arises in this way:
\begin{equation}
sc(-\eta_1+2\eta_3+4\eta_5+3\eta_6+2\eta_8)+4(c^2-s^2)(\eta_4+\eta_7)=0,
\label{eq:derivsub}
\end{equation}
where $c=\cos\varphi$ and $s=\sin\varphi$. Hence, altogether we
have four linear conditions for the nine $\eta_i$ coefficients
coming from Eqs.~(\ref{eq:eigenstate},\ref{eq:derivsub}). With
these four conditions, it is easy to see that the following linear
program provides the maximum quantum per local value for our W
state along with the measurement angles $\varphi_1=\varphi$ and
$\varphi_2=\varphi-\pi/2$:

\begin{equation}
\label{linprog}
\begin{aligned}
Q \equiv \max{  q}& \\
\text{subject to}\, \sum_{i=1}^9{E_{\lambda,i}b_i} &\le L \quad\forall \lambda\\
              \sum_{i=1}^9{R_{ji}b_i} - \eta_j &= 0\quad (j=1,\ldots,9)\\
              \sum_{i=1}^9{T_{ki}\eta_i} &= 0\quad (k=1,\ldots,4)\\
\end{aligned}
\end{equation}
where $q$ is the quantum value~(\ref{eq:expect}) to be maximized, $b_i$ and $\eta_i$ are
the variables to be determined, whereas $R$ is a $9\times 9$
matrix of coefficients coming from relations in
Eq.~(\ref{eq:eta(b)}) and $T$ is a $4\times 9$ matrix of
coefficients coming from the four
conditions~(\ref{eq:eigenstate},\ref{eq:derivsub}). We can fix
$L=1$ without loss of generality and $E_{\lambda,i}$ are the
symmetrized components of the local deterministic strategy
$\lambda$:
\begin{align}
E_{\lambda,1}&=A_1+B_1+C_1\nonumber\\
E_{\lambda,2}&=A_2+B_2+C_2\nonumber\\
E_{\lambda,3}&=A_1B_1+A_1C_1+B_1C_1\nonumber\\
E_{\lambda,4}&=A_1B_2+A_1C_2+B_1C_2+A_2B_1+A_2C_1+B_2C_1\nonumber\\
E_{\lambda,5}&=A_2B_2+A_2C_2+B_2C_2\nonumber\\
E_{\lambda,6}&=A_1B_1C_1\nonumber\\
E_{\lambda,7}&=A_1B_1C_2+A_1B_2C_1+A_2B_1C_1\nonumber\\
E_{\lambda,8}&=A_2B_2C_1+A_2B_1C_2+A_1B_2C_2\nonumber\\
E_{\lambda,9}&=A_2B_2C_2,
\end{align}
where each $A_i,B_i,C_i$, $i=1,2$ may take the values of $\pm 1$,
and each strategy $\lambda$ is characterized by a particular
choice for these values. In our particular case, this amounts to
$2^8=64$ strategies. However, due to permutational symmetry of the
Bell polynomial ${\cal B}$ some of the deterministic strategies
give the same value. In fact, it is enough to take
$(4\times5\times6)/(1\times2\times3)=20$ different strategies.

\begin{table}[t]
\caption{Coefficients of the three Bell inequalities ${\cal B}_1$,
${\cal B}_2$, ${\cal B}_3$ maximally violated by the W state
(corresponding to coefficients $b_1^i$,$b_2^i$,$b_3^i$,
respectively): ${\cal B}_1$ gives the largest quantum per local
value $Q/L=1.49177284$ by angle $\varphi=0.09275644\pi$, ${\cal
B}_2$ and ${\cal B}_3$ belong to the angle $\varphi=\pi/4$
providing the respective ratios $Q/L=964/(872-48\sqrt{2})\approx
1.19883$ and $Q/L=1.16666$.} \vskip 0.2truecm
\begin{tabular}{r|r r r}
\hline
$i$&$b_1^i\quad\quad$&$b_2^i\quad\quad$&$b_3^i\quad\quad$\\
\hline
1&-0.28155401&$336-160\sqrt2$&0\\
2& 0.03986104&$336-160\sqrt2$&0\\
3&-0.18252567&$-132-6\sqrt{2}$&-1\\
4&-0.18252567&$-304+30\sqrt{2}$&-2\\
5& 0.15080767&$-132-6\sqrt{2}$&-1\\
6&-0.47003882&$30+89\sqrt{2}$&$3/(2\sqrt2)$\\
7&-0.28751315&$102-83\sqrt{2}$&$-1/(2\sqrt2)$\\
8& 0.17656653&$102-83\sqrt{2}$&$-1/(2\sqrt2)$\\
9&-0.04204495&$30+89\sqrt{2}$&$3/(2\sqrt2)$\\
\hline
\end{tabular}
\label{table:bestsol}
\end{table}

\begin{figure}[t]
\vspace{0.05cm}
\includegraphics[angle=-90, width=\columnwidth]{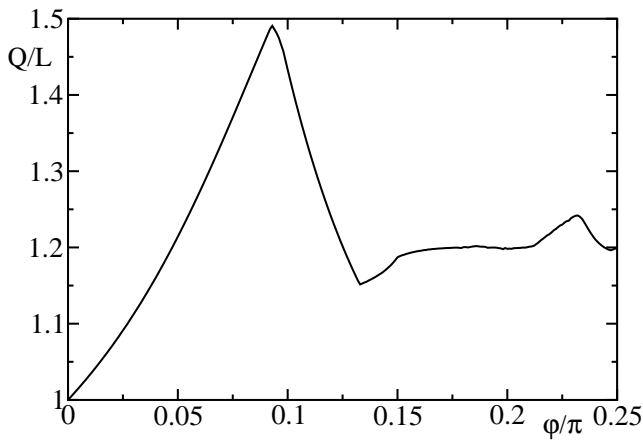} \caption{Dependence of the largest
quantum/local value on measurement angle $\varphi$ for a Bell
inequality violated maximally by the W state.} \label{fig:Qpcphi}
\end{figure}

We solved the above LP~(\ref{linprog}) by scanning through the
interval $\varphi=0\ldots\pi/4$. Fig.\ {\ref{fig:Qpcphi}} shows
the resulting $Q/L$ value as a function of $\varphi$. Notice that
according to the figure there is no appropriate solution at
$\varphi=0$. Incidentally, this implies that the W state with $Z$
and $X$ measurements cannot be self-tested: The set of
correlations arising from this particular state and measurements
is not unique.

We have chosen three particular Bell inequalities (denoted by
${\cal B}_1$, ${\cal B}_2$, and ${\cal B}_3$) according to the
relative measurement angle $\varphi$. The coefficients of the
respective Bell inequalities $b_{1,2,3}^i$, ($i=1,\ldots,9$) are
given in Table~\ref{table:bestsol}. (i) ${\cal B}_1$: the angle
$\varphi=0.09275644\pi$ which corresponds to the largest $Q/L$
ratio of 1.49177284. For this inequality the local bound is $L=1$.
(ii) ${\cal B}_2$: the angle $\varphi=\pi/4$, in which case the
Bell coefficients become symmetric under the exchange of the two
measurements $M_1$ and $M_2$. The classical limit is
$L=872-48\sqrt{2}$, while the quantum maximum is $Q=964$, giving
the ratio of $Q/L\approx 1.19883$. (iii) ${\cal B}_3$: the angle
$\varphi=\pi/4$ and we restrict ourselves to Bell inequalities
without marginals (that is $b_1=b_2=\eta_1=\eta_2=0$), in which
case we get a solution with the not much smaller $Q/L=7/6$ with
somewhat nicer looking coefficients presented in
Table~\ref{table:bestsol}. In that case, $\eta_3=-3$;
$\eta_5=\eta_8=1$, all other $\eta_i$ are zero, in which case
$L=6$ and $Q=7$. Note the values given at $\varphi=\pi/4$ are
exact. This can be checked by making use of the dual formulation
of the LP~(\ref{linprog}).

Let us stress that the constraints we have derived are only
necessary conditions for the W state to be the one which violates
the Bell inequality maximally. For the right solution the W state
must be the eigenstate belonging to the maximum eigenvalue, and
there must not exist another state with some different measurement
operators giving the same or larger violation. This extra
condition, for instance, is not guaranteed by our procedure.

We used see-saw method \cite{seesaw} in two-dimensional
component Hilbert spaces to test our conjecture. Let us note that
since the number of inputs and outputs of our inequalities is 2,
it is enough to verify the conjecture for d=2 \cite{lluis}. Any
other higher dimensional state can be decomposed as a direct sum
of $N$-qubit states. If all such states are unitarily equivalent
to the W state (and all measurement operators equal to X and Z),
we know that we can self-test W with that high dimensional state.

Running see-saw from independent random seeds many times,
we could recover the W state as the optimal state corresponding to
the reported maximal violations of the inequalities in
Table~\ref{table:bestsol}. This supports that the Bell
inequalities are good candidates for self-testing of the W state.
The drawback of the see-saw method, however, is that it is a
heuristic method and therefore it is not guaranteed to find the
solution (i.e. the specific state and measurements) corresponding
to maximal quantum violation. This limitation can be circumvented
by applying the Navascues-Pironio-Acin (NPA) method~\cite{npa},
which algorithmic process characterizes the quantum set from
outside without imposing dimensionality constraints. Using NPA
hierarchy on level~3 we find that our solution of W state along
with orthogonal measurements indeed saturates the upper bound
provided by the NPA method up to high numerical accuracy. However,
in order to prove conclusively that the maximal Bell violations of
Table~\ref{table:bestsol} are attained only by W states we will
make use of the SWAP method~\cite{swap} which gives us a powerful
numerical tool to estimate the distance of a produced state from
the W state in function of Bell violation. Incidentally, this
method originates in the NPA hierarchy.

\section{SWAP method and results}\label{swapmethod}

Here we just give the basic idea of the SWAP method and in the
further subsections we then give the results for self-testing of
different multipartite states. For a detailed explanation of the
method, we refer the reader to Ref.~\cite{swap}, which discusses
thoroughly the bipartite case but the generalization to more
parties is straightforward.

Suppose that we want to show that a multipartite state produced in
a Bell experiment is close to a desired state, which we denote by
$\bar{|\psi\rangle}$. The only information we have access to is
the experimental violation $Q$ of a given Bell inequality
$\cal{B}$. The SWAP method \cite{swap} combines (i) the idea of
swapping black boxes with trusted systems \cite{my} with (ii)
the semidefinite characterization of quantum correlations \`a la NPA~\cite{npa}.

(i) Let $\rho_{ABC}$ be the black-box system and let the trusted
auxiliary qubits $A',B',C'$ be prepared in the state $|0\rangle$.
Then some local unitaries $U_{AA'}$, $U_{BB'}$, $U_{CC'}$ are applied between the trusted
systems and their respective boxes, which operations leave the
trusted system in the state
\begin{equation}\label{rhoswap}
\rho_{swap}={\textrm{Tr}}_{ABC}(U\rho_{ABC}\otimes|000\rangle\langle000|_{A'B'C'}U^{\dagger}),
\end{equation}
where $U=U_{AA'}\otimes U_{BB'}\otimes U_{CC'}$. We want to choose $U$ such that the fidelity
\begin{equation}\label{fid}
F=\langle\bar{\psi}|\rho_{swap}|\bar{\psi}\rangle
\end{equation}
is as large as possible.

However, the virtual operation $U$ must be evaluated only from the
mere knowledge of statistical data (e.g. from the amount of a
Bell violation). At this point comes the NPA method to our help.

(ii) The crucial observation \cite{npa} is that, for an arbitrary
state $|\psi\rangle$ and set of operators $\{\hat M_i\}$, the
matrix $\Gamma$ with entries
$\Gamma_{ij}={\textrm{Tr}}(|\psi\rangle\langle\psi|\hat
M_i^{\dagger}\hat M_j)$ is positive semidefinite.

How does this help? For illustration, consider a
three-party situation, and let $S$ be a set of products of the
following operators $A_x,B_y,C_z$:
$S=\{(\one,A_1,A_2,A_1A_2,A_2A_1)\times(\one,B_1,B_2,B_1B_2,B_2B_1)\times(\one,C_1,C_2,C_1C_2,C_2C_1)$.
This set has $N=5\times5\times5=125$ components which we denote by
$M_i$, $i=1,\ldots,N$. According to the above remark, the $N$-dimensional $\Gamma$ matrix built up out of these operators must be
positive semidefinite. Moreover, some of the matrix elements are
equal or satisfy other constraints (for instance, all diagonal
entries have to be 1). Such constraints we collectively denote by
${\textrm{Tr}}(\alpha_i\Gamma)=\delta_i$, $i=1,\ldots,K$, where
$K$ is the number of constraints, and matrices $\alpha_i$ and
scalars $\delta_i$ are associated with the constraints. Finally,
noting that both the fidelity expression~(\ref{fid}) and the Bell
value are linear combinations of certain entries of the $\Gamma$
matrix, we obtain the following semidefinite programming (SDP)
\cite{BV} relaxation of the original problem:
\begin{equation}
\label{sdpprog}
\begin{aligned}
f = \min{  \textrm{Tr}(\tilde{F}\Gamma)} \\
\text{subject to}\, \Gamma&\ge 0\\
{\textrm{Tr}}(\alpha_i\Gamma)&=\delta_i\quad (i=1,\ldots,K)\\
               {\textrm{Tr}}(\tilde{B}\Gamma) &= Q,\\
\end{aligned}
\end{equation}
where $\tilde{B}$ is the matrix which contains our Bell inequality
in question and $\tilde{F}$ is the matrix encompassing the
device-independent fidelity expression. Matrices $\alpha_i$
contain linear constraints. By solving this program, which can be
done using standard SDP packages, we obtain a
lower bound $f$ on the true fidelity of the quantum state
$\rho_{swap}$ to a given reference state $|\bar{\psi}\rangle$.

Let us next summarize the computational resources used in solving
the SDP problem~(\ref{sdpprog}) above. In all studied cases we
used the MATLAB modeling language YALMIP \cite{yalmip}. For the
three-qubit computations, the size of the $\Gamma$ matrix is
$125\times 125$ and the number of constraints is $K=8604$. In this
case, we also increased the size of the $\Gamma$ matrix by
including in sequence $S$ the following third-order terms
$A_1A_2A_1$, $B_1B_2B_1$, $C_1C_2C_1$ (with $\Gamma$ matrix having
dimension $6^3=216$, and $K=24436$). However, to our surprise we did
not get any improvement over the previous results (the difference
in all values were in the range of $10^{-8}$, which is roughly the
precision of our SDP solver). In both cases, we used SeDuMi
\cite{sedumi} as a solver and solving the SDP for a single instance of Bell violation
took about 1 hour and 1 day, respectively, on a standard desktop PC.

As for the four-qubit computations, the size of the $\Gamma$
matrix is $625\times 625$ and the number of constraints is
$K=202186$. In this case, we had to use the SDPNAL solver
\cite{sdpnal}, which in spite of the large number of constraints
solved the SDP problem (for one instance of Bell violation) within half an hour.

\subsection{Self-testing of W state}\label{selftestW}

\begin{figure}[tl]
\vspace{0.05cm}
\includegraphics[angle=0, width=\columnwidth]{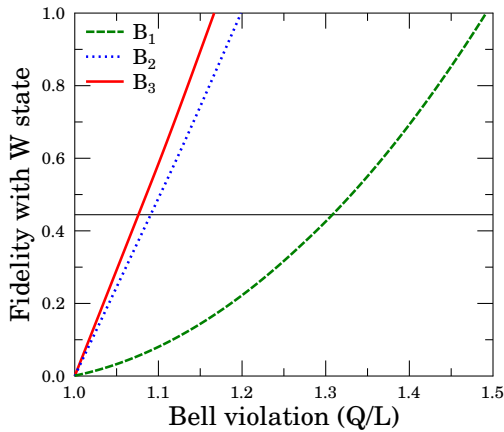} \caption{
Robust self-testing of the W state. Minimal fidelity with respect
to the ideal W state as a function of Bell violations $B_1,B_2$, and $B_3$.} \label{fig:W}
\end{figure}

We give below the details for the self-testing of the W state via
the SWAP method using the three Bell expressions
${\cal{B}}_{1,2,3}$ in Table~\ref{table:bestsol}. The lower bound
results for the fidelity $F$ are shown in Fig.~\ref{fig:W}.  These
curves can be directly used in Bell experiments to certify how
close a black box state is to a three-qubit W state. Noting that
by replacing the swaps in (\ref{rhoswap}) by identity operators
acting on trusted qubits which are initialized in some product
state guarantees that a fidelity of $4/9$ can be achieved with
respect to the W state (which value is independent of the Bell
violation). Hence, we expect that the curves provide useful
information only above this threshold (whose value of $4/9$ is
designated by solid black line).

Also note that, for the SWAP method to work, the optimal
measurement settings have to be the Pauli $Z$ and $X$, instead of
our rotated measurements, in which case we have to rotate our W
state correspondingly. Hence, the state we actually self-test is
$|\tilde{W}\rangle=U\otimes U\otimes U|W\rangle$ and the
corresponding measurements are $A_1=B_1=C_1=\hat Z$ and
$A_2=B_2=C_2=\hat X$, where $U=\cos(\pi/4-\varphi/2)\one -
i\sin(\pi/4 - \varphi/2)\hat Y$. Since this kind of local isometry
is part of the definition of self-testing, we can still identify
this state with the W state. Similar rotation tricks have been
applied to the GHZ and cluster states in the next subsections.

\subsection{Self-testing of GHZ states}\label{selftestGHZ}

\begin{figure}[tr]
\vspace{0.05cm}
\includegraphics[angle=0, width=\columnwidth]{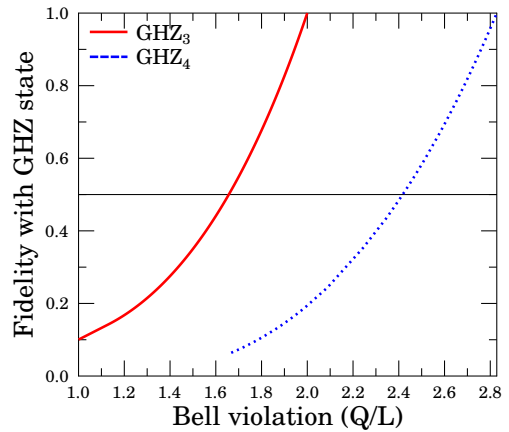} \caption{
Robust self-testing of GHZ states. Minimal fidelity
with respect to the ideal GHZ states as a function of
Bell violation (Mermin and MABK inequalities).}
\label{fig:GHZ}
\end{figure}

We perform robust self-testing for the (i) three-qubit GHZ
state~(\ref{ghzstate3}) using the Mermin-Bell
expression~(\ref{mermin3}) and for the (ii) four-qubit GHZ
state~(\ref{ghzstate4}) using the MABK-Bell
expression~(\ref{mermin4}). In both cases, the fidelity of $1/2$
can be attained with the $|000\rangle$ product state, hence the
figure gives useful information only above this threshold value
(presented with a black solid line). Please see Figure~\ref{fig:GHZ}.

In a recent experiment, DiCarlo et al.~\cite{dicarlo} use
superconducting circuits to implement the three-qubit GHZ state
with a fidelity of $87\pm 1\%$, as assessed via full state
tomography. DiCarlo et al. also evaluate the Mermin sum
(\ref{mermin3}), obtaining the value $Q=3.4\pm0.1$, or,
equivalently, $Q/L = (3.4\pm0.1)/2=1.7\pm 0.05$. For such a Bell
violation, the certified fidelity value is $F=57\%$, as can be
read off from solid curve~\ref{fig:GHZ}. This nicely demonstrates
the power of the device-independent approach. While our certified
fidelity is (obviously) below the one reported in
Ref.~\cite{dicarlo}, it has the advantage that it does not depend
on any details of the measurement devices used in the experiment.

\subsection{Self-testing of the cluster state}\label{selftestcluster}

The four-qubit linear cluster state \cite{cluster} to be used in our robust self-testing is
\begin{equation}\label{clusterstate}
|Cl\rangle = \frac{1}{2}(|0000\rangle + |0011\rangle +
|1100\rangle - |1111\rangle).
\end{equation}
Note that this state is not permutationally invariant.

We consider the Bell inequality that results when adding up the
inequalities defined by eq.~(26) and eq.~(27a) in T\'oth et al.~\cite{clusterNL1}:
\begin{align}
\label{tothetal}
&Toth\equiv A_1C_1D_2+A_2B_1C_2D_2+A_1C_2D_1-A_2B_1C_1D_1\nonumber\\
&+B_2C_1D_2+A_2B_1C_2D_2+B_2C_2D_1-A_2B_1C_1D_1\le4
\end{align}

\begin{figure}[tr]
\vspace{0.05cm}
\includegraphics[angle=0, width=\columnwidth]{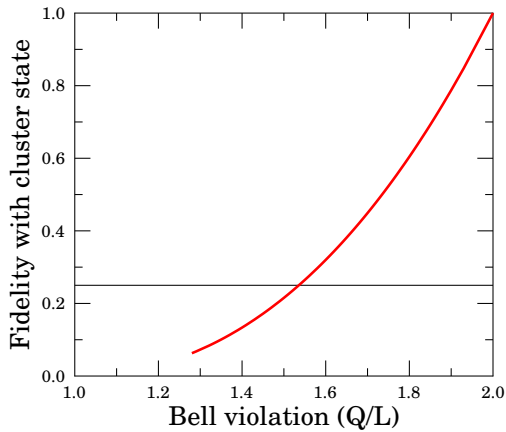} \caption{
Robust self-testing of the cluster state. Minimal fidelity with
respect to the ideal cluster state as a function of Bell violation
(T\'oth et al. inequality~(\ref{tothetal})).} \label{fig:CL}
\end{figure}

The T\'oth et al. Bell expression above can attain the algebraic
maximum of 8 with a cluster state. The respective settings are
$\hat Z$ and $\hat X$ up to local rotations. Hence, this
inequality is a good candidate for self-testing. The minimal
certified fidelity in function of the Bell
violation~(\ref{tothetal}) is shown in figure~\ref{fig:CL}. We
recall that the fidelity of $1/4$ can be attained with a product
state, hence the figure gives useful information only above this
threshold value (drawn in a black solid line).

The four-qubit cluster state (\ref{clusterstate}) has been
implemented with photons \cite{cluster_exp} and recently in a
system of trapped ions \cite{lanyon} as well. In the first case,
the two-setting Scarani et al.~inequality~\cite{scarani} was used
in a Bell experiment, for which the cluster state is not a unique
eigenstate of the Bell operator giving maximal violation. Hence,
it is not suitable for self-testing. In the second case, the
three-setting G\"uhne et al.~ inequality~\cite{guhne} was used in
the Bell test, in which case the cluster state is a unique
eigenstate of the Bell operator, hence suitable for self-testing.
Unfortunately, the computational resources required to
implement the swap method in the four-party/three-setting Bell
scenario are too demanding for a normal desktop.

\section{Conclusion}\label{conc}

In this paper, we have presented an efficient algorithm
based on linear programming to generate multipartite Bell
inequalities which are good candidates for self-testing of
permutationally invariant states. In combination with the SWAP
method \cite{swap}, the new inequalities and other famous Bell
functionals have allowed us to self-test the W state and other
notable multipartite states, such as the GHZ and cluster states.
Our main findings are summarized in
Figs.~\ref{fig:W},\ref{fig:GHZ},\ref{fig:CL}, which show how far
the black-box state is (in terms of the fidelity measure) from a
reference state for a given Bell violation. The presented lower
bounds for the fidelity are promising from an experimental point
of view, and, as we showed, some of them actually apply to recent
experiments.

We have some open questions. The computational effort of the
swap method for generic Bell inequalities scales badly with the
number of parties. Let us recall that for four parties the number
of SDP constraints are $\sim 2\times10^5$. However,
permutationally invariant Bell inequalities carry lots of
additional symmetries over generic Bell inequalities which might
be exploited to reduce the complexity of the SDP problem to be
solved. This simplification may allow the swap method to be
applied beyond four qubit-systems.

Self-testing of higher dimensional systems has already been
demonstrated through the example of the bipartite three-outcome CGLMP inequality
\cite{swap}. It would be challenging to self-test three-party
higher dimensional states as well, such as the fully
anti-symmetric state (also called Aharonov state used in the
Byzantine agreement problem \cite{aharonov}) or the generalized
three-qudit GHZ state $|\psi\rangle =
\sum_{i=0}^{d-1}{|i\rangle|i\rangle|i\rangle}/\sqrt{d}$ for
$d\ge3$.

Four-qubit (or even more complex) entangled states are routinely
generated and characterized in various types of systems, including
photons \cite{yao}, ions \cite{monz,w8qubit}, and superconducting
qubits \cite{dicarlo}. Due to the experimentally friendly nature
of the device-independent approach, we find it intriguing to
perform nonlocality experiments based on our Bell expressions in
Table~\ref{table:bestsol} and extract certified fidelity values
from our respective curves in Fig.~\ref{fig:W}.

As shown in Ref.~\cite{swap}, the SWAP method is also useful to
self-test measurement devices in the bipartite scenario. It would
be interesting to generalize our results concerning self-testing
of multipartite quantum states to the realm of self-testing
measurements in the multipartite scenario.

\section*{Acknowledgements}
K.F.P. and T.V. acknowledges financial support from the
T\'AMOP-4.2.2.C-11/1/KONV-2012-0001 project. The project has been
supported by the European Union, cofinanced by the European Social
Fund. T.V. acknowledges financial support from a J\'anos Bolyai
Grant of the Hungarian Academy of Sciences and the Hungarian
National Research Fund OTKA (PD101461). M.N. acknowledges the
European Commission (EC) STREP "RAQUEL", as well as the MINECO
project FIS2008-01236, with the support of FEDER funds.

\begin{appendix}

\section{Deriving an extra condition}\label{app}

Here it is shown that the mean value $\langle W|\hat{\cal
B}|W\rangle$ does not change in first order on small variations
around the measurement angles $\varphi_1$ and $\varphi_2$.
Let us take $\varphi_1=\varphi+\delta_1$ and $\varphi_2=\varphi-\pi/2+\delta_2$.
Then, neglecting second order terms, from Eq.\ (\ref{eq:M12phi12}) it follows:
\begin{align}
\hat M_1&=(c-s\delta_1)\hat Z+(s+c\delta_1)\hat X\nonumber\\
\hat M_2&=(s+c\delta_2)\hat Z-(c-s\delta_2)\hat X,
\label{eq:M12phi}
\end{align}
where $c\equiv\cos\varphi$ and $s\equiv\sin\varphi$.
By substituting these expressions into Eq.\ (\ref{eq:B_operators}) we get through straightforward calculation:
\begin{align}
\hat G_1=&(c-s\delta_1)\hat H_1+(s+c\delta_1)\hat H_2\nonumber\displaybreak[0]\\
\hat G_2=&(s+c\delta_2)\hat H_1-(c-s\delta_2)\hat H_2\nonumber\displaybreak[0]\\
\hat G_3=&(c^2-2sc\delta_1)\hat H_3+[sc+(c^2-s^2)\delta_1]\hat H_4+\nonumber\\
&(s^2+2sc\delta_1)\hat H_5\nonumber\displaybreak[0]\\
\hat G_4=&2(sc-s^2\delta_1+c^2\delta_2)\hat H_3+\nonumber\\
&[s^2-c^2+2sc(\delta_1+\delta_2)]\hat H_4-\nonumber\\
&2(sc+c^2\delta_1-s^2\delta_2)\hat H_5\displaybreak[0]\nonumber\displaybreak[0]\\
\hat G_5=&(s^2+2sc\delta_2)\hat H_3-[sc+(c^2-s^2)\delta_2]\hat H_4+\nonumber\\
&(c^2-2sc\delta_2)\hat H_5\displaybreak[0]\nonumber\displaybreak[0]\\
\hat G_6=&c^2(c-3s\delta_1)\hat H_6+c[sc+(c^2-2s^2)\delta_1]\hat H_7+\nonumber\\
&s[sc+(2c^2-s^2)\delta_1]\hat H_8+s^2(s+3c\delta_1)\hat H_9\nonumber\displaybreak[0]\\
\hat G_7=&3c(sc-2s^2\delta_1+c^2\delta_2)\hat H_6+\nonumber\\
&[c(2s^2-c^2)+3sc^2\delta_2+2s(2c^2-s^2)\delta_1]\hat H_7+\nonumber\\
&[s(s^2-2c^2)+3s^2c\delta_2+2c(2s^2-c^2)\delta_1]\hat H_8-\nonumber\\
&3s(sc+2c^2\delta_1-s^2\delta_2)\hat H_9\nonumber\displaybreak[0]\\
\hat G_8=&3s(sc-s^2\delta_1+2c^2\delta_2)\hat H_6+\nonumber\\
&[s(s^2-2c^2)+3s^2c\delta_1+2c(2s^2-c^2)\delta_2]\hat H_7+\nonumber\\
&[c(c^2-2s^2)-3sc^2\delta_1+2s(s^2-2c^2)\delta_2]\hat H_8+\nonumber\\
&3c(sc+c^2\delta_1-2s^2\delta_2)\hat H_9\nonumber\displaybreak[0]\\
\hat G_9=&s^2(s+3c\delta_2)\hat H_6-s[sc+(2c^2-s^2)\delta_2]\hat H_7+\nonumber\\
&c[sc+(c^2-2s^2)\delta_2]\hat H_8-c^2(c-3s\delta_2)\hat H_9.
\label{eq:B(H)}
\end{align}
By substituting these expressions into Eq.\ (\ref{eq:Bell-operator}) and comparing the result to
Eq.\ (\ref{eq:Bell-operatorH}) one can express the $\eta_i$ coefficients with $b_i$, and the
angles characterizing the measurement operators. Then by using Eq.\ (\ref{eq:Bell-operatorH})
one gets for the expectation value of the Bell-operator:
{\allowdisplaybreaks\begin{align}
&\langle W|\hat{\cal B}|W\rangle=\{(c-s\delta_1)b_1+(s+c\delta_2)b_2\}-\nonumber\\
&\{(c^2-2sc\delta_1)b_3+2(sc-s^2\delta_1+c^2\delta_2)b_4+\nonumber\\
&\vphantom{W}(s^2+2sc\delta_2)b_5\}+\nonumber\\
&2\{(s^2+2sc\delta_1)b_3-2(sc+c^2\delta_1-s^2\delta_2)b_4+\nonumber\\
&\vphantom{W}(c^2-2sc\delta_2)b_5\}-\nonumber\\
&\{c^2(c-3s\delta_1)b_6+3c(sc-2s^2\delta_1+c^2\delta_2)b_7+\nonumber\\
&\vphantom{W}3s(sc-s^2\delta_1+2c^2\delta_2)b_8+s^2(s+3c\delta_2)b_9\}+\nonumber\\
&2\{s[sc+(2c^2-s^2)\delta_1]b_6+\nonumber\\
&\vphantom{W}[s(s^2-2c^2)+3s^2c\delta_2+2c(2s^2-c^2)\delta_1]b_7+\nonumber\\
&\vphantom{W}[c(c^2-2s^2)-3sc^2\delta_1+2s(s^2-2c^2)\delta_2]b_8+\nonumber\\
&\vphantom{W}c[sc+(c^2-2s^2)\delta_2]b_9\}.
\label{eq:expect(b)}
\end{align}}
We must choose the coefficients such that the derivatives of the expression above in terms of
$\delta_1$ and $\delta_2$ are zero, that is:
\begin{align}
&-sb_1+6scb_3+2(s^2-2c^2)b_4+s(7c^2-2s^2)b_6+\nonumber\\
&\vphantom{-}2c(7s^2-2c^2)b_7+3s(s^2-2c^2)b_8=0\label{eq:deriv(b1)}\\
&-cb_2-6scb_5+2(2s^2-c^2)b_4+c(2c^2-7s^2)b_9+\nonumber\\
&\vphantom{-}2s(2s^2-7c^2)b_8+3c(2s^2-c^2)b_7=0.
\label{eq:deriv(b2)}
\end{align}
These are necessary conditions for $\varphi$ and $\varphi-\pi/2$ to be the optimal
measurement angles. They can also be expressed with the $\eta_i$ coefficients with this choice of
angles. We can get those by substituting Eqs.\ (\ref{eq:B(H)}) at $\delta_1=\delta_2=0$
into Eq.\ (\ref{eq:Bell-operator}) and comparing the result to
Eq.\ (\ref{eq:Bell-operatorH}):
{\allowdisplaybreaks\begin{align}
\eta_1=&cb_1+sb_2\nonumber\\
\eta_2=&sb_1-cb_2\nonumber\\
\eta_3=&c^2b_3+2scb_4+s^2b_5\nonumber\\
\eta_4=&scb_3-(c^2-s^2)b_4-scb_5\nonumber\\
\eta_5=&s^2b_3-2scb_4+c^2b_5\nonumber\\
\eta_6=&c^3b_6+3sc^2b_7+3s^2cb_8+s^3b_9\nonumber\\
\eta_7=&sc^2b_6-c(c^2-2s^2)b_7+s(s^2-2c^2)b_8-s^2cb_9\nonumber\\
\eta_8=&s^2cb_6+s(s^2-2c^2)b_7+c(c^2-2s^2)b_8+sc^2b_9\nonumber\\
\eta_9=&s^3b_6-3s^2cb_7+3sc^2b_8-c^3b_9, \label{eq:eta(b)}
\end{align}}
which can be written formally as $\eta_i=\sum_j{R_{ij}b_j}$,
$i=1,\ldots,9$.

It is easy to see from Eq.\ (\ref{eq:M12phi}) that if
$\delta_1=\delta_2=0$, the $(\hat M_1, \hat M_2)$ pair may be
expressed with $(\hat Z, \hat X)$ the same way than the other way
around. Therefore, Eqs.~(\ref{eq:eta(b)}) and the inverse
relationships has the same coefficients, i.e.
$b_i=\sum_j{R_{ij}\eta_j}$, $i=1,\ldots,9$.

Now let us add Eq.\ (\ref{eq:deriv(b1)}) to Eq.\
(\ref{eq:deriv(b2)}). Comparing the result to Eqs.\
(\ref{eq:eta(b)}) it is fairly easy to see that the result is:
\begin{equation}
-\eta_2+6\eta_4+7\eta_7-2\eta_9=0.
\label{eq:derivsum}
\end{equation}

However, if the $|W\rangle$ is an eigenstate of the Bell-operator, this relationship
is automatically fulfilled as the equation follows from  Eqs.~(\ref{eq:eigenstate}).
If we multiply Eq.~(\ref{eq:deriv(b1)}) by $c^2$ and Eq.~(\ref{eq:deriv(b2)}) by $s^2$,
subtract them from each other, and use $c^2+s^2=1$ several times,
a somewhat lengthier calculation does lead to an independent, fairly simple equation:
\begin{equation}
sc(-\eta_1+2\eta_3+4\eta_5+3\eta_6+2\eta_8)+4(c^2-s^2)(\eta_4+\eta_7)=0.
\label{eq:derivsub_appendix}
\end{equation}
This is the condition appearing in Eq.~(\ref{eq:derivsub}) in the
main text. We note that the derivation of
Eq.~(\ref{eq:derivsub_appendix}) and the spurious
Eq.~(\ref{eq:derivsum}) is not a crucial step. Instead of
Eq.~(\ref{eq:derivsub}), we could have taken both
Eqs.~(\ref{eq:deriv(b1)},\ref{eq:deriv(b2)}) directly as
constraints for the linear program.

\end{appendix}


\begin{thebibliography}{99}

\bibitem{rmp_horo} R. Horodecki, P. Horodecki, M. Horodecki, and K. Horodecki, Rev. Mod. Phys. {\bf 81}, 865 (2009).

\bibitem{review_toth} O. G\"uhne and G. T\'oth, Phys. Rep. {\bf 474}, 1 (2009).

\bibitem{rosset} D. Rosset, R. Ferretti-Sch\"obitz, J-D. Bancal, N. Gisin, and Y-C. Liang, Phys. Rev. A {\bf 86}, 062325 (2012).

\bibitem{photonic} G.~T\'oth et al., Phys. Rev. Lett. {\bf 105}, 250403 (2010);
 W. Wieczorek et al., Phys. Rev. Lett. {\bf 103}, 020504 (2009).

\bibitem{pistate} C. Schwemmer, G. T\'oth, A. Niggebaum, T. Moroder, D. Gross, O. G\"uhne, H. Weinfurter, arXiv:1401.7526 (2014).

\bibitem{monz} T. Monz et al., Phys. Rev. Lett. {\bf 106}, 130506 (2011).

\bibitem{lanyon}
B.P. Lanyon, M. Zwerger, P. Jurcevic, C. Hempel, W. D\"ur, H.J.
Briegel, R. Blatt, and C.F. Roos, Phys. Rev. Lett. {\bf 112},
100403 (2014).

\bibitem{gross} D. Gross, Y.-K. Liu, S. T. Flammia, S. Becker, and J.
Eisert, Phys. Rev. Lett. {\bf 105}, 150401 (2010).

\bibitem{mps} M. Cramer, M. B. Plenio, S. T. Flammia, R. Somma, D.
Gross, S. D. Bartlett, O. Landon-Cardinal, D. Poulin, and Y.-K.
Liu, Nat. Commun. {\bf 1}, 149 (2010).

\bibitem{review_scarani} V. Scarani, arXiv:1303.3081 (2013).

\bibitem{my} D. Mayers and A. Yao, Quant. Inf. Comput. {\bf 4}, 273 (2004).

\bibitem{selftest_ideal}
S.J. Summers, R.F. Werner, Commun. Math. Phys. {\bf 110}, 247
(1987); S. Popescu, D. Rohrlich, Phys. Lett. A {\bf 169}, 411
(1992); B.S. Tsirelson, Hadronic Journal Supplement {\bf 8}, 329
(1993).

\bibitem{selftest_nonideal}
C.-E. Bardyn, T.C.H. Liew, S. Massar, M. McKague, and V. Scarani,
Phys. Rev. A {\bf 80}, 062327 (2009); M. McKague, T.H. Yang, and
V. Scarani, J. Phys. A: Math. Theor. {\bf 45}, 455304 (2012); C.A.
Miller and Y. Shi, arXiv:1207.1819 (2012); B.W. Reichardt, F.
Unger and U. Vazirani, Nature {\bf 496}, 456 (2013); T.H. Yang and
M. Navascu\'es, Phys. Rev. A {\bf 8}, 050102(R) (2013).

\bibitem{swap} T.H. Yang, T. Vertesi, J.-D. Bancal, V. Scarani,
and M. Navascues, arXiv:1307.7053 (2013); arXiv:1406.7127 (2014).

\bibitem{chsh}
J.F. Clauser, M.A. Horne, A. Shimony, and R.A. Holt, Phys. Rev.
Lett. {\bf 23}, 880 (1969).

\bibitem{dur} W. Dur, G. Vidal, and J.I. Cirac, Phys. Rev. A {\bf 62}, 062314 (2000).

\bibitem{ghz} D.M. Greenberger, M.A. Horne, A. Zeilinger, Bells Theorem, Quantum Theory, and Conceptions of the Universe (ed. M. Kafatos, Kluwer Academic, Dordrecht, Holland, 1989), p. 69.

\bibitem{cluster} H.J. Briegel and R. Raussendorf, Phys. Rev. Lett. {\bf 86}, 910 (2001);
R. Raussendorf, H. J. Briegel, Phys. Rev. Lett. {\bf 86}, 5188 (2001).

\bibitem{w_exp}
M. Eibl, S. Gaertner, M. Bourennane, C. Kurtsiefer, M. Zukowski,
and H. Weinfurter, Phys. Rev. Lett. {\bf 90}, 200403 (2003).

\bibitem{ghz_exp}
J.-W.~Pan, D. Bouwmeester, M. Daniell, H. Weinfurter, and A.
Zeilinger, Nature {\bf 403}, 515 (2000); M. Eibl, N. Kiesel, M.
Bourennane, C. Kurtsiefer, and H. Weinfurter, Phys. Rev. Lett.
{\bf 92}, 077901 (2004); Z. Zhao, T. Yang, Y.-A. Chen, A.-N.
Zhang, M. Zukowski, and J.-W. Pan, Phys. Rev. Lett. {\bf 91},
180401 (2003).

\bibitem{cluster_exp}
P. Walther, M. Aspelmeyer, K. J. Resch, and A. Zeilinger, Phys.
Rev. Lett. {\bf 95}, 020403 (2005); N. Kiesel et al., Phys. Rev.
Lett. {\bf 95}, 210502 (2005).

\bibitem{merminpara} N. D. Mermin, Am. J. Phys. {\bf 58}, 731 (1990).

\bibitem{mckague} M. McKague, arXiv:1010.1989 (2010).

\bibitem{gme} J.-D. Bancal, N. Gisin, Y.-C. Liang, and S. Pironio, Phys. Rev. Lett. {\bf 106}, 250404 (2011);
K. F. Pal, T. Vertesi, Phys. Rev. A {\bf 83}, 062123 (2011).

\bibitem{gmeexp}  J. T. Barreiro, J-D. Bancal,  P. Schindler, D. Nigg,  M. Hennrich, T. Monz, N. Gisin, and R. Blatt, Nature Physics {\bf 9}, 559 (2013).

\bibitem{moroder} T. Moroder, J.-D. Bancal, Y.-C. Liang, M. Hofmann and O. G\"{u}hne, Phys. Rev. Lett. {\bf 111}, 030501 (2013).

\bibitem{neg} G. Vidal and R. F. Werner, Phys. Rev. A {\bf 65}, 032314 (2002).

\bibitem{structure} N. Brunner, J. Sharam, and T. V\'ertesi, Phys. Rev. Lett. {\bf 108}, 110501 (2012).

\bibitem{mirror_paper} X. Wu, Y. Cai, T. H. Yang, H. N. Le, J.-D. Bancal and V. Scarani,  arXiv:1407.5769.

\bibitem{bell}
J. S. Bell, Physics 1, 195 (1964); N. Brunner, D. Cavalcanti, S.
Pironio, V. Scarani and S. Wehner, Rev. Mod. Phys. {\bf 86}, 419
(2014).

\bibitem{mermin} N. D. Mermin, Phys. Rev. Lett. {\bf 65}, 1838 (1990).

\bibitem{MABK} M. Ardehali, Phys. Rev. A {\bf 46}, 5375 (1992); A.V. Belinskii and D.N. Klyshko, Phys. Usp. {\bf 36}, 653 (1993).

\bibitem{PI_Bell} J.-D. Bancal, N. Gisin, and S. Pironio, J. Phys. A: Math. Theor. {\bf 43}, 385303 (2010); N. Brunner, J. Sharam, and T. Vertesi, Phys. Rev. Lett. {\bf 108}, 110501 (2012); J. Tura, A. B. Sainz, T. Vertesi, A. Acin, M. Lewenstein, and R. Augusiak, arXiv:1312.0265 (2013).

\bibitem{dicke} R.H. Dicke, Phys. Rev. {\bf 93}, 99 (1954).

\bibitem{seesaw}
R.F. Werner and M.M. Wolf, Quantum Inf. Comput. {\bf 1}, 1 (2001);
K.F. P\'al and T. V\'ertesi, Phys. Rev. A {\bf 82}, 022116 (2010).

\bibitem{lluis} Ll. Masanes, arXiv:quant-ph/0512100 (2005).

\bibitem{npa} M. Navascu\'es, S. Pironio, and A. Ac\'in, Phys. Rev. Lett. {\bf 98}, 010401 (2007);
M. Navascu\'es, S. Pironio and A. Ac\'in, New J. Phys. {\bf 10},
073013 (2008).

\bibitem{BV} S.P. Boyd and L. Vandenberghe, {\it Convex Optimization},
Cambridge University Press, 2004.

\bibitem{yalmip} J. L\"ofberg, YALMIP: A Toolbox for Modeling and Optimization in MATLAB, {\it Proceedings of the CACSD Conference}
(Taipei, Taiwan, 2004).

\bibitem{sedumi}
J.F. Sturm, {\it Using SeDuMi 1.02, a MATLAB toolbox for
optimization over symmetric cones}, Optimization Methods and
Software {\bf 11} 625 (1999).

\bibitem{sdpnal} X. Zhao, D. Sun, and K.-C. Toh, SIAM J. Optim. {\bf 20}, 1737 (2010).

\bibitem{dicarlo} L. DiCarlo et al., Nature {\bf 467}, 574 (2010).

\bibitem{clusterNL1} G. Toth, O. Guehne, H. J. Briegel, Phs. Rev. A {\bf 73}, 022303 (2006).

\bibitem{scarani}
V. Scarani, A. Ac\'in, E. Schenck, and M. Aspelmeyer, Phys. Rev. A
{\bf 71}, 042325 (2005).

\bibitem{guhne}
O. G\"uhne, G. T\'oth, P. Hyllus, and H.J. Briegel, Phys. Rev.
Lett. {\bf 95}, 120405 (2005).

\bibitem{aharonov} M. Fitzi, N. Gisin, U. Maurer, Phys. Rev. Lett. {\bf 87}, 217901 (2001).

\bibitem{yao} X.-C. Yao et al., Nat. Phot. {\bf 6}, 225 (2012).

\bibitem{w8qubit} H. H\"{a}ffner, W. H\"{a}nsel, C. Roos, J. Benhelm, et al., Nature
{\bf 438}, 643 (2005).






\end{thebibliography}
\end{document}